\documentclass[11pt]{article}
\usepackage{amssymb}
\begin{document}
\title{Generalization of solutions of the Jacobi PDEs \\
associated to time reparametrizations of Poisson systems}
\author{Benito Hern\'{a}ndez--Bermejo $^a$}
\date{}
\maketitle

\vspace{-1.2cm}

\begin{center}
{\em Departamento de Matem\'{a}tica Aplicada. Universidad Rey Juan Carlos.} \\ 
{\em Escuela Superior de Ciencias Experimentales y Tecnolog\'{\i}a. Edificio Departamental II.} \\ 
{\em Calle Tulip\'{a}n S/N. 28933--M\'{o}stoles--Madrid. Spain.} 
\end{center}

\mbox{}

\noindent \rule{15.7cm}{0.01in}
\noindent{\bf Abstract} \vspace{2mm}

The determination of solutions of the Jacobi partial differential equations (PDEs) for 
finite-dimensional Poisson systems is considered. In particular, a novel procedure for the construction of solution families is developed. Such procedure is based on the use of time reparametrizations preserving the existence of a Poisson structure. As a result, a method which is valid for arbitrary values of the dimension and the rank of the Poisson structure under consideration is obtained. In this article two main families of time reparametrizations of this kind are characterized. In addition, these results lead to a novel application which is also developed, namely the global and constructive determination of the Darboux canonical form for Poisson systems of arbitrary dimension and rank two, thus improving the local result provided by Darboux' theorem for such case. 

\mbox{}


\noindent \rule{15.7cm}{0.01in}

\vfill

\footnoterule
\noindent $^a$ Telephone: (+34) 91 488 73 91. Fax: (+34) 91 488 73 38. \newline 
\mbox{} \hspace{0.05cm} E-mail: benito.hernandez@urjc.es

\pagebreak
\begin{flushleft}
{\bf 1. Introduction}
\end{flushleft}

The search, classification and analysis of solutions of the Jacobi partial differential equations (PDEs) \cite{lic1,olv1,wei1}, has deserved an important attention during the last decades \cite{agz1,bhq1,cgr1,dddh},\cite{ghha}-\cite{byv1},\cite{byv4,hoj2,auo1,lyx1,tym1}. Such equations are given by the system of nonlinear coupled PDEs
\begin{equation}
     \label{jac}
     \sum_{l=1}^n ( J_{li} \partial_l J_{jk} + J_{lj} \partial_l J_{ki} + 
     J_{lk} \partial_l J_{ij} ) = 0 \:\; , \;\:\;\: i,j,k=1, \ldots ,n
\end{equation}
where $ \partial_j \equiv \partial / \partial x_j$, and at the same time the so-called structure functions $J_{ij}(x_1, \ldots ,x_n) \equiv J_{ij}(x)$ must also verify the additional skew-symmetry conditions: 
\begin{equation}
     \label{sksym}
     J_{ij} =  - J_{ji} \:\; , \;\:\;\: i,j=1, \ldots ,n
\end{equation}
Equations (\ref{jac}-\ref{sksym}) are defined in a domain (open and connected set) $\Omega \subset \mathbb{R}^n$ in such a way that the $C^{\infty}( \Omega )$ functions $J_{ij}(x)$ conform in $\Omega$ the entries of an $n \times n$ structure matrix ${\cal J}(x)$ which can be degenerate in rank. Such rank will be denoted by $r$ in what follows. 

The applied interest of this problem is due to its key role in the framework of Poisson systems. Recall that a finite-dimensional dynamical system defined in an domain $\Omega \subset \mathbb{R}^n$ is said to be Poisson (or to define a Poisson structure in $\Omega$) if it can be written \cite{olv1} in terms of a smooth set of ODEs of the form: 
\begin{equation}
    \label{nham}
    \frac{\mbox{d}x_i}{{\mbox{d}t}} = \sum_{j=1}^n J_{ij}(x) \partial _j H(x) 
	\; , \;\:\; i = 1, \ldots , n, 
\end{equation} 
or briefly $\dot{x}= {\cal J}(x) \cdot \nabla H(x)$, where ${\cal J}(x) \equiv (J_{ij}(x))$ is a structure matrix, and function $H(x)$, which is by construction a first integral, plays the role of Hamiltonian. 

Finite-dimensional Poisson systems (e.g. see \cite{olv1} and references therein for an overview and a historical discussion) are ubiquitous in most fields of applied mathematics and related areas such as physics and theoretical biology, for instance in mechanics 
\cite{cchh,ghha,gyn1,lzhh,sim1}, control theory \cite{bpr1}, electromagnetism \cite{cyl1,dddh}, plasma physics \cite{pyj1}, 
optics \cite{dht3,hwo1}, population dynamics \cite{byv2,byv5,nut2,nut1,pla1}, 
dynamical systems theory \cite{bhq1,cgr1,cchh,hoj2,lzhh,mag1}, etc. 
In fact, describing a given dynamical system in terms of a Poisson structure allows the obtainment of a wide range of information which may be in the form of perturbative solutions \cite{cyl1}, invariants \cite{dddh,lzhh,tym1}, bifurcation properties and characterization of chaotic behaviour \cite{dht3,lzhh,pyj1}, efficient numerical integration \cite{jay1}, integrability results \cite{lzhh,mag1,olv1}, reductions 
\cite{almr,dddh,dht3,bs2},\cite{bs4}-\cite{bsnf},\cite{byv2,byv4}, as well as algorithms for stability analysis \cite{bpmm,byv5,hyct,opbr,sim1}, to cite a sample. 

Moreover, there are additional fundamental reasons justifying the importance and interest of Poisson systems. One is that they provide a generalization of classical Hamiltonian systems, allowing not only for odd-dimensional vector fields, but also because a structure matrix verifying (\ref{jac}-\ref{sksym}) admits a great diversity of forms apart from the classical (constant) symplectic matrix. Actually, Poisson systems are a generalization of classical Hamiltonian systems on which a (typically) noncanonical Poisson bracket is defined, namely:
\begin{equation}
\label{z1cp}
	\{ f(x),g(x) \} = \sum _{i,j=1}^n \partial_i f(x) J_{ij}(x) \partial_j g(x)
\end{equation}
for every pair of smooth functions $f(x)$ and $g(x)$. The possible rank degeneracy of the structure matrix ${\cal J}$ implies that a certain class of first integrals ($D(x)$ in what follows) termed Casimir or distinguished invariants exist. There is no analog in the framework of classical Hamiltonian theory for such constants of motion, which have the property of having zero bracket in the sense of (\ref{z1cp}) with all smooth functions defined in $\Omega$, namely: $\{ D(x),f(x) \} =0$ for every $f(x)$. It can be seen that this implies that Casimir invariants are the solution set of the system of coupled PDEs: ${\cal J}(x) \cdot \nabla D(x) =0$. The determination of Casimir invariants and their use in order to carry out a reduction (local, in principle) is the cornerstone of the (at least local) dynamical equivalence between Poisson systems and classical Hamiltonian systems, as stated by Darboux' theorem \cite{olv1}: 

\mbox{}

\noindent {\bf Theorem 1.1 (Darboux).} {\em Consider an $n$-dimensional Poisson system defined in a domain $\Omega \subset \mathbb{R}^n$ for which the rank of the structure matrix has constant value $r$ everywhere in $\Omega$. Then at each point of $\Omega$ there exist local coordinates $(p_1, \ldots ,p_{r/2},q_1, \ldots , q_{r/2},z_1, \ldots , z_{n-r})$ in terms of which the equations of motion become:
\[
	\dot{q}_i = \frac{\partial H}{\partial p_i} \;\: , \;\:\:
	\dot{p}_i = - \frac{\partial H}{\partial q_i} \;\: , \;\:\: i=1, \ldots ,r/2 
\]
\[
	\dot{z}_j = 0 \;\: , \;\:\: j=1 , \ldots , n-r
\]}

\vspace{-3mm}
\mbox{}

This justifies that Poisson systems can be regarded, to a large extent, as a generalization of classical Hamiltonian systems. This connection is an additional and important advantage of Poisson systems, as far as it accounts for the potential transfer of results and techniques from classical Hamiltonian theory once a given system has been recognized as a Poisson one and the Darboux canonical form has been constructed, specially if this can be done globally in the domain of interest \cite{almr,dddh,dht3,bs2},\cite{bs4}-\cite{bsnf},\cite{byv2,byv4}. 

In addition, the problem of describing a given vector field not explicitly written in the form (\ref{nham}) in terms of a Poisson structure is a fundamental question in this context, which still remains as an open issue \cite{bhq1,cgr1,byv1,hoj2,per1}. This is a nontrivial decomposition to which some efforts have been devoted in past years following different approaches. The source of the difficulty is twofold: First, a known constant of motion of the system able to play the role of the Hamiltonian is required. And second, it is necessary to find a suitable structure matrix for the vector field. Consequently, finding a solution of the Jacobi identities (\ref{jac}) complying also with conditions (\ref{sksym}) is unavoidable. 

Together, the previous features explain the attention deserved in the literature by the obtainment, classification and analysis of skew-symmetric solutions of the Jacobi equations. 
Among the very diverse procedures used for the identification of new solutions of 
(\ref{jac}-\ref{sksym}), one possibility consists of the use of a previously known structure matrix which is taken as starting point. In other words one structure matrix, assumed to be known, is the basis used in the search of more general solutions. This kind of strategy often has the advantage of simplifying notably the Jacobi equations. In particular, approaches of this type have been successfully used in different contexts \cite{bou1,dddh,bs1,byv2}. In the present work an investigation of this kind is carried out, as it will be described in detail in the next section. For the moment, two properties of the forthcoming analysis are worth being emphasized. In first place, the following results provide a generalization of solutions which is valid for arbitrary values of the dimension $n$ and the rank $r$ of the Poisson structure under consideration ---an uncommon feature in the analysis of the Jacobi PDEs. In second term, the generalization methodology considered here is dynamically meaningful as far as it is naturally associated to the analysis of time reparametrizations of Poisson systems. Equivalently, the investigation to be performed can be formulated in terms of the conditions to be verified in arbitrary dimension $n$ by a structure matrix and a time reparametrization if the Poisson structure is to be preserved after the time reparametrization is applied to the associated Poisson system. In this article two main solution families of time reparametrizations of this kind are characterized. In addition, these results lead to a novel application that is also developed. Such application is the constructive and global determination of the Darboux canonical form for Poisson systems of arbitrary dimension and rank two. It is worth recalling that the global determination of the Darboux coordinates in arbitrary-dimensional Poisson systems is typically a nontrivial task only known for a few classes of Poisson structures \cite{bsn1,bsnf,byv2,byv4,pyj1,wei1}. Accordingly, the present application leads to an additional global construction of this kind, thus improving the local scope of Darboux' theorem for such case.

The structure of the article is the following. In Section 2 a detailed description of the problem under consideration is presented. Sections 3 and 4 are respectively devoted to the characterization of two solution families of time reparametrizations which preserve the existence of a Poisson structure for generic Poisson systems. The work concludes in Section 5 which develops an application to the global construction of the Darboux canonical form for 
rank-two Poisson structures of arbitrary dimension. 

\mbox{}

\begin{flushleft}
{\bf 2. Formulation of the problem}
\end{flushleft}

Provided ${\cal J}(x)$ is an $n$-dimensional structure matrix of constant rank in a domain $\Omega \subset \mathbb{R}^n$, we shall now consider the following problem: given an arbitrary function $\eta (x): \Omega \rightarrow \mathbb{R}$ which is smooth in $\Omega$ and does not vanish in $\Omega$, we shall investigate the conditions such that the product $\eta (x) {\cal J}(x)$ is also a structure matrix. 

The naturalness of this question should be clear because of the close relationship of this issue with the problem of determining whether or not a specific new-time transformation (or NTT, also called time reparametrization in the literature) applied to a Poisson system preserves the existence of a Poisson structure. In order to explain this, let us begin with the following:

\mbox{}

\noindent{\bf Definition 2.1.} Given a smooth dynamical system 
\[
\frac{\mbox{d}x}{\mbox{d}t} = f(x)
\] 
with $x \in \mathbb{R}^n$, a new-time transformation (or NTT) is a reparametrization of the time variable of the form 
\begin{equation}
\label{ndntt}
	\mbox{\rm d}\tau = \frac{1}{\eta (x)}\mbox{\rm d}t
\end{equation}
where $t$ is the initial time variable, $\tau$ is the new time and $\eta (x) : \Omega \subset 
\mathbb{R}^n \rightarrow \mathbb{R}$ is a smooth function in $\Omega$ which does not vanish in $\Omega$. 

\mbox{}

Thus, given a Poisson system (\ref{nham}) defined in $\Omega$, the time reparametrization 
(\ref{ndntt}) leads from (\ref{nham}) to the system (not necessarily of Poisson type) of the form:
\begin{equation}
\label{psrtnd}
	\frac{\mbox{\rm d}x}{\mbox{\rm d} \tau} = \eta {\cal J} \cdot \nabla H
\end{equation}
It is already known in the literature \cite{gyn1,bs2} that the new vector field (\ref{psrtnd}) obtained from (\ref{nham}) after a general time reparametrization is always a Poisson system in the cases of dimensions $n=2$ and $n=3$. On the contrary, this is not necessarily the situation for dimensions $n \geq 4$, which are then our main subject in what follows. In addition, apart from being a natural problem in this framework, the study of time reparametrizations is interesting because sometimes \cite{bs2,bs4,bs3,bsnf} (but not always 
\cite{dddh,bsn1,byv2,byv4,pyj1}) their use is necessary in order to achieve the Darboux canonical form for given families of Poisson systems. As we shall see in Section 5, the investigation of time reparametrizations provides criteria and results of applied interest for the global determination of the Darboux canonical form.

In this context, it is also convenient to give a brief explanation about the condition $\eta (x) \neq 0$ in $\Omega$ just introduced. Of course, it would be mathematically acceptable to investigate the conditions such that $\eta (x) {\cal J}(x)$ is a structure matrix provided $ {\cal J}(x)$ is, with the only requirement of a smooth $\eta (x)$. However, according to Definition 2.1 such problem could not be assimilated to the use of time reparametrizations, which is of central interest in the present context. Additionally, a second key reason for choosing a nonvanishing function $\eta (x)$ is derived from the fact that if Rank(${\cal J}(x)$) is constant in $\Omega$, then the rank of $\eta (x) {\cal J}(x)$ will be also constant in $\Omega$. The interest in this constancy is of course the applicability of Darboux' theorem, also of importance in this work. 

The following definition is natural for the problem considered:

\mbox{}

\noindent {\bf Definition 2.2.} Let ${\cal J}(x)$ be an $n \times n$ structure matrix defined everywhere in a domain $\Omega \subset \mathbb{R}^n$ and of constant rank in $\Omega$, and let $\eta (x) : \Omega \rightarrow \mathbb{R}$ be a smooth function which does not vanish in $\Omega$ and such that $\eta (x) {\cal J}(x)$ is also a structure matrix defined everywhere in $\Omega$. Then, the function $\eta (x)$ will be called a reparametrization factor for ${\cal J}(x)$ in $\Omega$. 

\mbox{}

In connection with the previous definition, it is necessary to provide the following result:

\mbox{}

\noindent {\bf Proposition 2.1.} {\em Let ${\cal J}(x)$ be an $n \times n$ structure matrix defined everywhere in a domain $\Omega \subset \mathbb{R}^n$ and of constant rank in $\Omega$, and let $\eta (x)$ be a reparametrization factor for ${\cal J}(x)$ in $\Omega$. Then: 
\begin{description}
\item[\mbox{\rm {\em (a)}}] Function $D(x)$ is a Casimir invariant of $\eta (x) {\cal J}(x)$ in $\Omega$ if and only if it is a Casimir invariant of ${\cal J}(x)$ in $\Omega$. 
\item[\mbox{\rm {\em (b)}}] If a Poisson system having the structure matrix ${\cal J}(x)$ can be reduced globally and diffeomorphically in $\Omega$ to the Darboux canonical form, then every Poisson system having the structure matrix $\eta (x) {\cal J}(x)$ can also be reduced globally and diffeomorphically in $\Omega$ to the Darboux canonical form.
\end{description}
}

\mbox{}

\noindent {\bf Proof.} The proof of (a) is clear since Casimir invariants are the solution set of the system of PDEs given by ${\cal J} \cdot \nabla D =0$. Regarding (b), for the reduction of $\eta (x) {\cal J}(x)$ it suffices to perform a preliminary time reparametrization 
$\mbox{\rm d}\tau = \eta (x) \mbox{\rm d}t$, where as usual $t$ is the initial time variable, and $\tau$ is the new time. The outcome is thus a Poisson system with structure matrix ${\cal J}(x)$ and time variable $\tau$. The rest of the global reduction then follows the diffeomorphic steps known by hypothesis for ${\cal J}(x)$. \hfill \mbox{\bf {\small Q.E.D.}}

\mbox{}

The previous proposition thus implies that the identification of a reparametrization factor for a family of structure matrices immediately generalizes such family, while the operational framework provided by the knowledge of the Casimir invariants and the global Darboux reduction for the initial solution family is preserved in the generalization. In addition, the investigation of reparametrization factors is relevant as far as it aims at characterizing those Poisson structures that are not destroyed by (certain, at least) time reparametrizations. 

With regard to the problem formulation, let us recall that the Jacobi PDEs (\ref{jac})
vanish identically if $i,j,k$ are not all different, as it can be easily verified. Thus, for convenience, in what follows we shall sometimes make use of equations (\ref{jac}) with $i,j,k=1, \ldots ,n$ together with the additional conditions $i \neq j$, $i \neq k$ and $j \neq k$. As indicated earlier, we assume that an $n$-dimensional structure matrix ${\cal J}(x)$ is defined in a domain $\Omega \subset \mathbb{R}^n$, together with a function $\eta (x)$ which is $C^{\infty}(\Omega)$ and does not vanish in $\Omega$. If we substitute the product $\eta (x) {\cal J} (x)$ in equations (\ref{jac}) we arrive to the conditions: 
\[
     \eta \sum_{l=1}^n ( J_{il} \partial_l J_{jk} + J_{kl} \partial_l J_{ij} + 
	J_{jl} \partial_l J_{ki} ) + 
	\sum_{l=1}^n (J_{il} J_{jk} + J_{kl} J_{ij} + J_{jl} J_{ki}) \partial _l \eta 
	= 0 \:\; , \;\:\;\: i,j,k=1, \ldots ,n
\]
Since ${\cal J}$ is by hypothesis a structure matrix, this leads to:
\begin{equation}
\label{rtfjac}
      \sum_{l=1}^n (J_{il} J_{jk} + J_{kl} J_{ij} + J_{jl} J_{ki}) \partial _l \eta 
	= 0 \:\; , \;\:\;\: i,j,k=1, \ldots ,n
\end{equation}
Consistently, we see that equations (\ref{rtfjac}) vanish if two or three of the indexes $i$, $j$ and $k$ take the same value. Moreover, identities (\ref{rtfjac}) also vanish identically if one of such indexes coincides with $l$, even in the case in which $i$, $j$ and $k$ are all different. Accordingly, equations (\ref{rtfjac}) can be equivalently expressed as:
\begin{equation}
\label{rtf2jac}
	\sum_{ \stackrel{\scriptstyle l=1}{\scriptstyle l \neq i,j,k} }^n
	(J_{il} J_{jk} + J_{kl} J_{ij} + J_{jl} J_{ki}) \partial _l \eta 
	= 0 \:\; , \;\:\;\: \left\{ \begin{array}{l} i,j,k=1, \ldots ,n \\ 
	i \neq j; \;\: i \neq k; \;\: j \neq k \end{array} \right.
\end{equation}
In the forthcoming developments, either form (\ref{rtfjac}) or (\ref{rtf2jac}) will be preferred according to convenience. Notice that the outcome of the {\em ansatz\/} $\eta (x) {\cal J}(x)$ is a new problem in which now only one unknown function $\eta (x)$ exists. Therefore, equations (\ref{rtfjac}) or (\ref{rtf2jac}) constitute a set of linear PDEs for a single dependent variable $\eta (x)$. These features imply a significant simplification of the problem. Note also that $\eta (x) = c \neq 0$, with $c \in \mathbb{R}$ being an arbitrary constant, is always a solution. This trivial result will become a particular case of the first solution family of reparametrization factors to be determined in brief. 

In what follows we shall provide two solution families relative to problem (\ref{rtfjac}) or (\ref{rtf2jac}). This is the purpose of the next two sections. 

\mbox{}

\begin{flushleft}
{\bf 3. First family of reparametrization factor solutions}
\end{flushleft}

The result corresponding to a first family of solutions of equations (\ref{rtfjac}) is described in the following: 

\mbox{}

\noindent {\bf Theorem 3.1.} {\em Let ${\cal J}(x)$ be an $n \times n$ structure matrix of constant rank everywhere in a domain $\Omega \subset \mathbb{R}^n$, and let $D(x)$ be a Casimir invariant of ${\cal J}(x)$ globally defined in $\Omega$. Then $D(x) {\cal J}(x)$ is a structure matrix everywhere in $\Omega$.
}

\mbox{}

\noindent {\bf Proof.} Let us consider the problem equations in the form (\ref{rtfjac}). Such identities can be written in the following way: 
\begin{equation}
\label{rtf1}
	J_{jk} \sum_{l=1}^n J_{il} \partial _l \eta + 
	J_{ij} \sum_{l=1}^n J_{kl} \partial _l \eta +
	J_{ki} \sum_{l=1}^n J_{jl} \partial _l \eta = 0 \:\; , \;\:\;\: i,j,k=1, \ldots ,n
\end{equation}
Thus, equations (\ref{rtf1}) can be expressed as:
\begin{equation}
\label{rtf12}
	J_{jk} ({\cal J} \cdot \nabla \eta )_i + 
	J_{ij} ({\cal J} \cdot \nabla \eta )_k +
	J_{ki} ({\cal J} \cdot \nabla \eta )_j = 0 \:\; , \;\:\;\: i,j,k=1, \ldots ,n
\end{equation}
Consequently, if $\eta (x)$ is a Casimir invariant, equations (\ref{rtf12}) are identically satisfied, as far as Casimir functions constitute the solution set of the system ${\cal J}(x) \cdot \nabla D(x) =0$. \hfill \mbox{\bf {\small Q.E.D.}}

\mbox{}

Theorem 3.1 has a direct consequence: 

\mbox{}

\noindent {\bf Corollary 3.1.} {\em Let ${\cal J}(x)$ be an $n \times n$ structure matrix of constant rank $r$ in a domain $\Omega \subset \mathbb{R}^n$, having $(n-r)$ functionally independent Casimir invariants globally defined in $\Omega$. Then there are $(n-r)$ functionally independent reparametrization factors for ${\cal J}(x)$ globally defined in $\Omega$, and every nonvanishing $C^{\infty}(\mathbb{R}^{n-r})$ function of them is also a reparametrization factor for ${\cal J}(x)$ everywhere in $\Omega$. 
}

\mbox{}

\noindent {\bf Proof.} It is sufficient to make use of the following two remarks: in first place, every $C^{\infty}$ function of one or more Casimir invariants is also a Casimir invariant; and secondly, as a consequence of the previous statement, a Casimir invariant $D(x)$ which is vanishing somewhere in a given domain $\Omega$ can be trivially replaced by a nonvanishing one functionally dependent on it. \hfill \mbox{\bf {\small Q.E.D.}}

\mbox{}

The previous results also allow regarding as a particular case the fact (already mentioned) that constants are always solutions of equations (\ref{rtfjac}), just as a consequence that constants are (trivial) Casimir invariants of every structure matrix. Constant reparametrization factors are thus always present, even in the symplectic case ($r=n$). On the other hand, if the rank is lower than the dimension ($r<n$) then the number of nonconstant reparametrization factors is infinity. We shall turn back to these issues in the next section.

There is an alternative perspective that shows the naturalness of the result in Theorem 3.1. For this, consider a Poisson system $\dot{x}= {\cal J}(x) \cdot \nabla H(x)$. If we rescale the Hamiltonian as $H^*(x) = \eta (x) H(x)$ with $\eta (x)$ being a Casimir invariant, then the new system remains as a Poisson one, namely $\dot{x}= {\cal J}(x) \cdot \nabla [\eta (x)H(x)]$. However, this implies that: 
\[
\dot{x} = {\cal J}(x) \cdot \nabla [ \eta (x)H(x)] = {\cal J}(x) \cdot [ \eta (x) \nabla H(x) + H(x) \nabla \eta (x)] = \eta (x) {\cal J}(x) \cdot \nabla H(x)
\]
And therefore it is clear that such rescaling of the Hamiltonian (which is equivalent to a rescaling of the structure matrix) must preserve the existence of a Poisson structure. 

The family of reparametrization factors just characterized corresponds to a sufficient (but not necessary) condition for the verification of equations (\ref{rtfjac}). A natural question is if additional solutions exist. The answer is positive, as the next section describes.

\mbox{}

\begin{flushleft}
{\bf 4. Second family of reparametrization factor solutions}
\end{flushleft}

Let us focus again on the problem of searching reparametrization factors, this time making use of the equations in the form (\ref{rtf2jac}). Obviously, a sufficient condition (different from the one previously considered in Theorem 3.1) for the verification of (\ref{rtf2jac}) is that:
\begin{equation}
\label{rtf3jac}
	J_{il} J_{jk} + J_{kl} J_{ij} + J_{jl} J_{ki} = 0 \:\; , \;\:\;\: 
	\left\{ \begin{array}{l} i,j,k,l=1, \ldots ,n \\ 
	i \neq j,k,l ; \hspace{2mm} j \neq k,l ; \hspace{2mm} k \neq l
 \end{array} \right.
\end{equation}
An interesting aspect of the conditions (\ref{rtf3jac}) is that they are merely algebraic, which is a remarkable simplification of the initial PDE problem. If (\ref{rtf3jac}) is verified, then every $C^{\infty}$ and nonvanishing function $\eta (x)$ will be a valid reparametrization factor. The investigation of this possibility is the subject of the next theorem, which is the main result of this section: 

\mbox{}

\noindent {\bf Theorem 4.1.} {\em Let ${\cal J}(x)$ be an $n \times n$ structure matrix defined in a domain $\Omega \subset \mathbb{R}^n$ and of constant rank $r$ everywhere in 
$\Omega$. Then the product $\eta (x) {\cal J}(x)$ is a structure matrix in $\Omega$ for every $C^{\infty}(\Omega)$ function $\eta (x)$ if and only if $ r \leq 2$. 
}

\mbox{}

\noindent {\bf Proof.} Every implication will be demonstrated separately.

In one sense, let us first demonstrate that if Rank(${\cal J}$)$\leq 2$, then the product by every $C^{\infty}$ function $\eta$ preserves the property of being a structure matrix. For this, consider the following submatrix of ${\cal J}$, which is obtained after deleting all its rows and columns different from those at the positions $i,j,k$ and $l$ (with $i,j,k,l$ all different):
\begin{equation}
\label{mpls}
	{\cal J}^{[ijkl]} = \left( \begin{array}{cccc} 
	   0 & J_{ij} & J_{ik} & J_{il} \\ -J_{ij} & 0 & J_{jk} & J_{jl} \\
	-J_{ik} & -J_{jk} & 0 & J_{kl}  \\ -J_{il} & -J_{jl} & -J_{kl} & 0 
	\end{array} \right)
\end{equation}
If Rank(${\cal J}$)$\leq 2$, then it must be $\mid \! {\cal J}^{[ijkl]} \! \mid = 0$ in 
(\ref{mpls}) for all possible values of the four indexes $i,j,k,l$. But notice that in fact it is:
\[
	\mid \! {\cal J}^{[ijkl]} \! \mid = (J_{il}J_{jk}+J_{kl}J_{ij}+J_{jl}J_{ki})^2
\]
Consequently, identities (\ref{rtf3jac}) are verified and the proof in this sense is already accomplished. 

Conversely, let us demonstrate that if the product by every $C^{\infty}$ function $\eta$ preserves the character of structure matrix, then Rank(${\cal J}$)$\leq 2$. For convenience, we shall equivalently prove  that if Rank(${\cal J}$)$\geq 4$, then the product by every possible $C^{\infty}$ function $\eta$ does not always preserve the property of being a structure matrix. For this, we shall consider a given point $x_0 \in \Omega$, and let $J_{ij}(x_0) \equiv a_{ij}$ for all $i,j = 1, \ldots ,n$. The first part of the following reasoning is close to the one employed for the construction of the normal form for skew-symmetric matrices. In first place, let us assume without loss of generality that $a_{12} \neq 0$. If this is not the case, it is always possible to place another nonzero element in the position $(1,2)$: let $a_{ij} \neq 0$, then we can permute the first and second rows with the $i$-th and $j$-th rows, respectively, and later the first and second columns of the resulting matrix can also be permuted with the $i$-th and $j$-th columns, respectively. Therefore, independently of the value of $a_{12}$ the outcome is the following skew-symmetric matrix
\begin{equation}
\label{jpio}
	S^*_{x_0}= \left( \begin{array}{ccccc}
	0 & a_{\pi_1 \pi_2} & a_{\pi_1 \pi_3} & \ldots & a_{\pi_1 \pi_n} \\
	a_{\pi_2 \pi_1} & 0 & a_{\pi_2 \pi_3} & \ldots & a_{\pi_2 \pi_n} \\
	\vdots & \vdots & \vdots & \mbox{} & \vdots \\
	a_{\pi_n \pi_1} & a_{\pi_n \pi_2} & a_{\pi_n \pi_3} & \ldots & 0 
	\end{array} \right) \equiv
	\left( \begin{array}{cccc}
	0 & a_{\pi_1 \pi_2} & \vline & E_{2 \times (n-2)} \\
	- a_{\pi_1 \pi_2} & 0 & \vline & \mbox{} \\ \hline
	E_{(n-2) \times 2} & \mbox{} & \vline & E_{(n-2) \times (n-2)}
	\end{array} \right) 
\end{equation}
where $( \pi_1, \ldots , \pi_n )$ is a permutation of $(1, \ldots ,n)$: if $a_{12} \neq 0$, then such permutation is the identical one; and if $a_{12} =0 $, the permutation is given by $\pi _1 =i$, $\pi _i =1$, $\pi _2 =j$, $\pi _j =2$, and $\pi _k =k$ for every $k$ different from $1,2,i$ and $j$. Thus matrix (\ref{jpio}) is our starting point in either case, with $a_{\pi_1 \pi_2} \neq 0$. In the right-hand side of (\ref{jpio}), the letter $E$ denotes three submatrices of the sizes indicated by their respective subindexes. Since row and column elementary operations do not alter the rank of a matrix, we can make use of them in order to transform (\ref{jpio}) into the skew-symmetric matrix:
\begin{equation}
\label{jpio2}
	S_{x_0}^{**}= \left( \begin{array}{cccc}
	0 & a_{\pi_1 \pi_2} & \vline & \mathbb{O}_{2 \times (n-2)} \\
	- a_{\pi_1 \pi_2} & 0 & \vline & \mbox{} \\ \hline
	\mathbb{O}_{(n-2) \times 2} & \mbox{} & \vline & \tilde{E}_{(n-2) \times (n-2)}
	\end{array} \right) 
\end{equation}
In equation (\ref{jpio2}) and in what follows, $\mathbb{O}_{p \times q}$ denotes the $p \times q$ null submatrix. In addition, in submatrix $\tilde{E}_{(n-2) \times (n-2)}$ of (\ref{jpio2}) we now have the entries, 
\[
	\tilde{E}_{(n-2) \times (n-2)} = \left( \begin{array}{cccc}
	0 & \tilde{a}_{\pi _3 \pi _4} & \ldots & \tilde{a}_{\pi _3 \pi _n} \\
	\tilde{a}_{\pi _4 \pi _3} & 0 & \ldots & \tilde{a}_{\pi _4 \pi _n} \\
	\vdots & \vdots & \mbox{} & \vdots \\
	\tilde{a}_{\pi _n \pi _3} & \tilde{a}_{\pi _n \pi _4} & \ldots & 0 
	\end{array} \right)
\]
where it is:
\begin{equation}
\label{asubpis}
	\tilde{a}_{\pi _k \pi _l} = a_{\pi _k \pi _l} + \frac{1}{a_{\pi _1 \pi _2}} 
	(a_{\pi _1 \pi _l}a_{\pi _2 \pi _k} - a_{\pi _1 \pi _k}a_{\pi _2 \pi _l})
	\:\; , \;\:\;\: k,l=3, \ldots ,n
\end{equation}
At this stage, since Rank($S_{x_0}^{**}$)$\geq 4$, there must be a nonzero element in $\tilde{E}_{(n-2) \times (n-2)}$: if $\tilde{a}_{\pi _3 \pi _4} \neq 0$, then we do not need to perform any changes for what is to follow. On the contrary, if $\tilde{a}_{\pi _3 \pi _4} = 0$ we can again permute rows and columns in such a way that the position $(3,4)$ is occupied by a nonzero entry 
$\tilde{a}_{\pi _k \pi _l}$ from $\tilde{E}_{(n-2) \times (n-2)}$ (with both $\pi _k$ and $\pi _l$ different from $\pi _1$ and $\pi _2$), the resulting matrix being also skew-symmetric. Consequently, we can assume without loss of generality that it is $\tilde{a}_{\pi _3 \pi _4} \neq 0$. Then, from matrix $S_{x_0}^{**}$ in (\ref{jpio2}) we can pick out the following submatrix composed by the intersection of the first four rows and columns:
\begin{equation}
\label{jpio3}
	\tilde{S}_{x_0}^{[1234]}= \left( \begin{array}{ccccc}
	0 & a_{\pi_1 \pi_2} & \vline & 0 & 0 \\
	- a_{\pi_1 \pi_2} & 0 & \vline& 0 & 0 \\ \hline
	0 & 0 & \vline & 0 & \tilde{a}_{\pi _3 \pi _4} \\
	0 & 0 & \vline & - \tilde{a}_{\pi _3 \pi _4} & 0 
	\end{array} \right) 
\end{equation}
with both $a_{\pi_1 \pi_2} \neq 0$ and $\tilde{a}_{\pi _3 \pi _4} \neq 0$, as indicated. The determinant of $\tilde{S}_{x_0}^{[1234]}$ in (\ref{jpio3}) is $\mid \! \tilde{S}_{x_0}^{[1234]} \! \mid = (a_{\pi_1 \pi_2} \tilde{a}_{\pi _3 \pi _4})^2 \neq 0$. Now without loss of generality and for the sake of clarity, let us assume $\pi _i = i$ for all $i=1, \ldots ,4$. From (\ref{asubpis}) we thus have that: 
\begin{equation}
\label{eaqo}
	(a_{12} \tilde{a}_{34})^2 = \left[ a_{12} \left( a_{34}+ \frac{1}{a_{12}}(a_{14}a_{23} - 
	a_{13}a_{24}) \right) \right]^2 \neq 0
\end{equation}
Equation (\ref{eaqo}) immediately implies that:
\begin{equation}
\label{eaqo2}
	J_{12}(x_0) J_{34}(x_0)+ J_{14}(x_0)J_{23}(x_0) + J_{31}(x_0)J_{24}(x_0) \neq 0
\end{equation}
Let us investigate the implications of (\ref{eaqo2}) in equations (\ref{rtf2jac}). For this 
we may consider, for instance, the equation in (\ref{rtf2jac}) corresponding to the choice $i=1$, $j=2$ and $k=3$. Such equation takes the form:
\begin{equation}
\label{eaqo3}
	(J_{14}J_{23}+ J_{34}J_{12}+ J_{24}J_{31})\partial_4 \eta + 
	\sum_{l=5}^n(J_{1l}J_{23}+ J_{3l}J_{12}+ J_{2l}J_{31}) \partial_l \eta =0
\end{equation}
Now two cases must be distinguished, namely $n=4$ and $n \geq 5$: 

\begin{description}
\item[{\bf {\em Case I:}}] $n=4$. We proceed by means of two auxiliary lemmas:

\mbox{}

\noindent {\bf Lemma 4.1.} {\em Let ${\cal J}(x) \equiv (J_{ij}(x))$ be an $n \times n$ 
skew-symmetric matrix defined in a domain $\Omega \subset \mathbb{R}^n$. Then, for every $x \in \Omega$ the quantities
\[
	\Xi _{ijkl}(x) \equiv J_{il}(x) J_{jk}(x)+ J_{kl}(x)J_{ij}(x) + J_{jl}(x)J_{ki}(x)
	\:\; , \;\:\;\: i,j,k,l=1, \ldots ,n
\]
are completely skew-symmetric in all the subindexes $(i,j,k,l)$.
}

\mbox{}

\noindent {\bf Proof of Lemma 4.1.} The result can be verified by direct evaluation of the index skew-symmetry properties. \hfill 
\mbox{\bf {\small Q.E.D.}} 

\mbox{}

The second lemma required now is:

\mbox{}

\noindent {\bf Lemma 4.2.} {\em Let ${\cal J}(x)$ be a $4 \times 4$ structure matrix defined in a domain $\Omega \subset \mathbb{R}^4$ and such that Rank(${\cal J}$)$\, =4$ everywhere in $\Omega$. Then the only possible reparametrization factors allowed for ${\cal J}(x)$ in $\Omega$ are the constant ones.
}

\mbox{}

\noindent {\bf Proof of Lemma 4.2.} Now equations (\ref{rtf2jac}) amount to:
\begin{equation}
\label{aqua1}
	(J_{il} J_{jk} + J_{kl} J_{ij} + J_{jl} J_{ki}) \partial _l \eta 
	= 0 
\end{equation}
where in (\ref{aqua1}) the indexes $(i,j,k,l)$ may be every possible permutation of $(1,2,3,4)$. Due to the skew-symmetry property demonstrated in Lemma 4.1, the number of independent equations in (\ref{aqua1}) is actually four: 
\begin{equation}
\label{aqua2}
	\left\{ \begin{array}{lcl} 
	(i,j,k,l)=(1,2,3,4) & \Rightarrow & 
	(J_{14} J_{23} + J_{34} J_{12} + J_{24} J_{31}) \partial _4 \eta = 0 \\
	(i,j,k,l)=(1,2,4,3) & \Rightarrow & 
	(J_{13} J_{24} + J_{43} J_{12} + J_{23} J_{41}) \partial _3 \eta = 0 \\
	(i,j,k,l)=(1,3,4,2) & \Rightarrow & 
	(J_{12} J_{34} + J_{42} J_{13} + J_{32} J_{41}) \partial _2 \eta = 0 \\
	(i,j,k,l)=(2,3,4,1) & \Rightarrow & 
	(J_{21} J_{34} + J_{41} J_{23} + J_{31} J_{42}) \partial _1 \eta = 0 
	\end{array} \right.
\end{equation}
In addition, if ${\cal J}$ is a regular $4 \times 4$ skew-symmetric matrix, its determinant is: 
\begin{equation}
\label{aqua3}
	\mid \! {\cal J} \! \mid = (J_{12} J_{34} + J_{31} J_{24} + J_{14} J_{23})^2 \neq 0
\end{equation}
Hypothesis (\ref{aqua3}) implies that equations (\ref{aqua2}) are actually simplified to $\partial _l \eta =0$ for all $l=1, \ldots ,4$, namely $\eta$ is a constant. Lemma 4.2 is thus proven. \hfill \mbox{\bf {\small Q.E.D.}} 

\mbox{}

Therefore $\eta (x)$ cannot be an arbitrary function when $n=4$ and Case I is demonstrated. Let us now turn to the second possibility considered:

\mbox{}

\item[{\bf {\em Case II:}}] $n \geq 5$. Notice now that equation (\ref{eaqo3}) is valid, in particular, at $x_0 \in \Omega$. Assume, for instance, that a function $\eta (x)$ is chosen in such a way that $\partial _4 \eta \neq 0$ at $x_0$. Then, equation (\ref{eaqo2}) implies that it is not possible at the same time to make the choice $\partial _l \eta = 0$ at $x_0$ for all $l \geq 5$. Consequently, function $\eta (x)$ cannot be arbitrary in the complementary case $n \geq 5$. This demonstrates Case II.
\end{description}

The proof of Theorem 4.1 is thus complete. \hfill \mbox{\bf {\small Q.E.D.}}

\mbox{}

The results provided in the framework of this second family of reparametrization factors now investigated, can be complemented by means of an additional result, which actually generalizes Lemma 4.2: 

\mbox{}

\noindent {\bf Theorem 4.2.} {\em Let ${\cal J}(x)$ be an $n \times n$ structure matrix ($n \geq 4$) defined in a domain $\Omega \subset \mathbb{R}^n$ and such that Rank(${\cal J}$) $=n$ everywhere in $\Omega$. Then the only possible reparametrization factors allowed for ${\cal J}(x)$ in $\Omega$ are the constant ones.
}

\mbox{}

\noindent {\bf Proof.} We begin with an auxiliary result: 

\mbox{}

\noindent {\bf Lemma 4.3.} {\em Consider the structure matrix
\[
	{\cal S}_n \equiv
	\left( \begin{array}{cc} 0 & 1 \\ -1 & 0 \end{array} \right) 
	\overbrace{ \oplus \ldots \oplus }^{n/2} 
	\left( \begin{array}{cc} 0 & 1 \\ -1 & 0 \end{array} \right) 
\]
with $n \geq 4$ an even integer. Let $\Omega \subset \mathbb{R}^n$ be a domain. Then the only possible reparametrization factors allowed for ${\cal S}_n$ in $\Omega$ are the constant ones.
}

\mbox{}

\noindent {\bf Proof of Lemma 4.3.} Let us consider four different cases for the entries of ${\cal S}_n$:

\begin{description}
\item[{\bf {\em Case I.}}] Let $i$ be odd, with $1 \leq i \leq (n-3)$. Now we choose indexes $(i,j,k)=(i,i+1,i+2)$. Then from equations (\ref{rtfjac}) we obtain:
\begin{equation}
\label{rtfjaci1}
      \sum_{l=1}^n (J_{il} J_{jk} + J_{kl} J_{ij} + J_{jl} J_{ki}) \partial _l \eta = 
	\sum_{l=1}^n J_{i+2,l} \partial _l \eta = J_{i+2,i+3} \partial _{i+3} \eta =
	\partial _{i+3} \eta = 0 
\end{equation}
and consequently (\ref{rtfjaci1}) implies $\partial _l \eta =0$ for $l = i+3 = 4, 6, \ldots n$.

\item[{\bf {\em Case II.}}] Now let $(i,j,k)=(1,3,4)$. Again from (\ref{rtfjac}) we are led to:
\[
      \sum_{l=1}^n (J_{il} J_{jk} + J_{kl} J_{ij} + J_{jl} J_{ki}) \partial _l \eta = 
	\sum_{l=1}^n J_{1l} \partial _l \eta = J_{12} \partial _2 \eta = \partial _2 \eta = 0 
\]

\item[{\bf {\em Case III.}}] This time we choose even values of $i$, with $2 \leq i \leq (n-2)$. Then, with indexes $(i,j,k)=(i,i+1,i+2)$ from equations (\ref{rtfjac}) we now have:
\begin{equation}
\label{rtfjaci3}
      \sum_{l=1}^n (J_{il} J_{jk} + J_{kl} J_{ij} + J_{jl} J_{ki}) \partial _l \eta = 
	\sum_{l=1}^n J_{il} \partial _l \eta = J_{i,i-1} \partial _{i-1} \eta =
	- \partial _{i-1} \eta = 0 
\end{equation}
and thus (\ref{rtfjaci3}) leads to $\partial _l \eta =0$ for $l = i-1 = 1, 3, \ldots ,(n-3)$.

\item[{\bf {\em Case IV.}}] Finally, let $(i,j,k)=(n-3,n-2,n)$. Therefore (\ref{rtfjac}) implies:
\[
      \sum_{l=1}^n (J_{il} J_{jk} + J_{kl} J_{ij} + J_{jl} J_{ki}) \partial _l \eta = 
	\sum_{l=1}^n J_{nl} \partial _l \eta = J_{n,n-1} \partial _{n-1} \eta = - \partial _{n-1} 	\eta = 0 
\]
\end{description}

Together, Cases I-IV provide the result stated in Lemma 4.3. \hfill \mbox{\bf {\small Q.E.D.}} 

\mbox{}

Let us now continue the main proof. For this, it is worth noticing that after a general smooth change of variables $y \equiv y(x)$ transforming a structure matrix ${\cal J}(x)$ into a new one ${\cal J}^*(y)$, every reparametrization factor $\eta (x)$ is converted into $\eta ^*(y) = \eta (x(y))$. To see this, it suffices to recall the general transformation rule for structure matrices subjected to smooth coordinate changes $y \equiv y(x)$:
\begin{equation}
\label{aquajdf}
      J^*_{ij}(y) = \sum_{k,l=1}^n \frac{\partial y_i}{\partial x_k} J_{kl}(x) 
	\frac{\partial y_j}{\partial x_l} \:\; , \;\:\;\: i,j = 1 , \ldots , n
\end{equation}
Clearly, according to (\ref{aquajdf}) the transformation of $\eta (x) {\cal J}(x)$ leads to $\eta ^*(y) {\cal J}^*(y)$, with $\eta ^*(y)=\eta (x(y))$, as indicated. Now let $x_0 \in \Omega$ be a point, and consider the value of the matrix at that point, namely ${\cal J}(x_0)$. It is well-known that there exists a regular matrix $E_{x_0}$ such that $E_{x_0} \cdot {\cal J}(x_0) \cdot E_{x_0}^T = {\cal S}_n$. On the basis of this relationship, we perform on ${\cal J}(x)$ the change of variables, diffeomorphic in $\mathbb{R}^n$, given by $y = E_{x_0} \cdot x$. According to (\ref{aquajdf}), the outcome is evidently ${\cal J}^*(y)= E_{x_0} \cdot {\cal J}(x(y)) \cdot E^T_{x_0}$. Let $y_0 \equiv 
E_{x_0} \cdot x_0$. Thus, in particular we have that ${\cal J}^*(y_0)= {\cal S}_n$. Consider then equation (\ref{rtfjac}) for the reparametrization factor in the new variables $y$: 
\begin{equation}
\label{rneg}
      \sum_{l=1}^n [ J^*_{il}(y) J^*_{jk}(y) + J^*_{kl}(y) J^*_{ij}(y) + 
	J^*_{jl}(y) J^*_{ki}(y) ] 
	\partial _{y_l} \eta ^*(y) = 0 \:\; , \;\:\;\: i,j,k=1, \ldots ,n
\end{equation}
Since equations (\ref{rneg}) are valid everywhere in $\Omega ^* = y( \Omega )$, they are valid in particular in $y_0 \in \Omega ^*$, namely:
\begin{equation}
\label{rneg2}
      \sum_{l=1}^n [J^*_{il}(y_0) J^*_{jk}(y_0) + J^*_{kl}(y_0) J^*_{ij}(y_0) + 
	J^*_{jl}(y_0) J^*_{ki}(y_0)] ( \left. \partial _{y_l} \eta ^*(y) \right| _{y_0}) = 0 
	\:\; , \;\:\;\: i,j,k=1, \ldots ,n
\end{equation}
Given that ${\cal J}^*(y_0)= {\cal S}_n$, as indicated, the analysis provided in Lemma 4.3 is immediately applicable to equations (\ref{rneg2}). Consequently we find that: 
\begin{equation}
\label{dety0}
	\left. \frac{\partial \eta ^*(y)}{\partial y_i} \right| _{y_0} = 0 
	\:\; , \;\:\;\: i = 1, \ldots ,n
\end{equation}
Taking into account that it is $\eta ^*(y)= \eta (x(y))$, or equivalently that $\eta (x)= \eta ^*(y(x))$, an application of the chain rule combined with (\ref{dety0}) shows that: 
\begin{equation}
\label{dety1}
	\left. \frac{\partial \eta (x)}{\partial x_i} \right| _{x_0} = 
	\left. \frac{\partial \eta ^*(y(x))}{\partial x_i} \right| _{x_0} = 
	\sum _{j=1}^n \left( \left. \frac{\partial \eta ^*(y)}{\partial y_j} \right| _{y_0} 	\right) \left( \left. \frac{\partial y_j}{\partial x_i} \right| _{x_0} \right) = 0 
	\:\; , \;\:\;\: i=1, \ldots ,n
\end{equation}
Since the analysis leading to (\ref{dety1}) can be carried out for every point $x_0 \in \Omega$, we conclude that actually it is $\partial _{x_i} \eta (x) =0$ everywhere in $\Omega$ for all $i=1, \ldots ,n$, namely $\eta (x)$ is in fact a constant. The proof of Theorem 4.2 is complete. \hfill \mbox{\bf {\small Q.E.D.}} 

\mbox{}

Of course, in Theorem 4.2 the maximal rank condition Rank(${\cal J}$)$\, =n$ implies that we are dealing with even values of the dimension $n$. In spite of being a somehow exclusive result, such theorem complements the previous contributions for the characterization of reparametrization factors.

In the next section the goal will be to provide a novel application of the results just developed.  

\mbox{}

\begin{flushleft}
{\bf 5. Application: global Darboux reduction for Poisson structures of rank two}
\end{flushleft}

An applied consequence of the study of time reparametrizations for Poisson systems (and in particular of the second family of reparametrization factors, characterized in Section 4) is the possibility of constructing the global Darboux reduction for Poisson systems having structure matrices of rank two and arbitrary dimension. This improves the scope of Darboux' theorem for such kind of systems. The result is given in the next theorem:

\mbox{}

\noindent {\bf Theorem 5.1.} {\em Let $\Omega \subset \mathbb{R}^n$ be a domain ($n \geq 2$) where is defined a Poisson system 
\[
	\frac{\mbox{\rm d}x}{\mbox{\rm d}t} = {\cal J}(x) \cdot \nabla H (x)
\]
having an $n \times n$ structure matrix ${\cal J}(x) \equiv (J_{ij}(x))$, and such that Rank(${\cal J}$)$\, =2$ everywhere in $\Omega$. Let $(D_3(x), \ldots , D_n(x))$ be a complete set of independent Casimir invariants of ${\cal J}(x)$ in $\Omega$. In addition, let $(d_1(x), d_2(x))$ be two arbitrary $C^{\infty}(\Omega)$ functions such that the transformation
\begin{equation}
\label{jr2tdf}
	\left\{ \begin{array}{cclcl}
	y_i & = & d_i(x) & \mbox{} , \mbox{} \:\; & i = 1,2 \\
	y_j & = & D_j(x) & \mbox{} , \mbox{} \:\; & j = 3, \ldots ,n 
	\end{array}  \right.
\end{equation} 
is one-to-one everywhere in $\Omega$ and its Jacobian matrix $M$ verifies:
\begin{equation}
\label{rr2j1}
	\mid \! M \! \mid = 
	\left| \frac{\partial (d_1(x), d_2(x), D_3(x), \ldots , D_n(x))}{\partial(x_1, 
	\ldots ,x_n)} \right| \neq 0  \:\; , \;\:\;\: \mbox{{\em for all}} \:\; x \in \Omega
\end{equation}
Then such Poisson system can be reduced globally in $\Omega$ to an one degree of freedom Hamiltonian system and the Darboux canonical form is accomplished globally and diffeomorphically in $\Omega$ in the new coordinate system $(y_1, \ldots ,y_n)$ and the new time $\tau$, where $(y_1, \ldots ,y_n)$ are given by transformation (\ref{jr2tdf}) which is a diffeomorphism globally defined in $\Omega$; while the new time $\tau$ is defined by the time reparametrization:
\begin{equation}
\label{jr2darbntt}
	\mbox{\rm d} \tau = \left( \left. \{ d_1(x), d_2(x) \}_{\cal J} \right) \right| _{x(y)} 
	\mbox{\rm d} t = 
	\left[ \left. (\nabla_x d_1(x))^T \cdot {\cal J}(x) \cdot ( \nabla_x d_2(x)) 
	\right] \right| _{x(y)} \mbox{\rm d} t \equiv \eta (y) \mbox{\rm d} t
\end{equation}
}

\mbox{}

\noindent{\bf Proof.} The constancy of Rank(${\cal J}$) implies that Darboux' theorem is applicable. In column matrix notation for the gradients, the Jacobian matrix of (\ref{jr2tdf}) can be written as:
\begin{equation}
\label{rr2j2}
	M \equiv 
	\frac{\partial (d_1(x), d_2(x), D_3(x), \ldots , D_n(x))}{\partial(x_1, \ldots ,x_n)} = 	\left( \nabla_x d_1(x) \:\;\:\; \nabla_x d_2(x) \:\;\:\; \nabla_x D_3(x) \:\;\:\; \ldots 
	\:\;\:\; \nabla_x D_n(x) \right) ^T
\end{equation}
Note in particular that functions $d_1(x)$ and $d_2(x)$ cannot be Casimir invariants because they are functionally independent of a complete set of independent Casimir functions. It is well-known that, according to (\ref{aquajdf}), the effect of (\ref{jr2tdf}) is to transform ${\cal J}(x)$ into a new structure matrix ${\cal J}^*(y) = M \cdot {\cal J} \cdot M^T$. With the help of (\ref{rr2j2}) we find, still in column matrix notation:
\begin{equation}
\label{rr2je1}
	{\cal J}^* = M \cdot \left( [{\cal J} \cdot \nabla_x d_1(x)] \;\:\;\:\;\: 
	[{\cal J} \cdot \nabla_x d_2(x)] \;\:\;\:\;\: 
	\mathbb{O}_{n \times 1} \;\:\;\:\;\: \ldots \;\:\;\:\;\: \mathbb{O}_{n \times 1} 	\right) 
\end{equation}
Using in (\ref{rr2je1}) the fact that for any pair of matrices $A$ and $B$ that can be multiplied, we can write $A \cdot B = (B^T \cdot A^T)^T$, we immediately find that (\ref{rr2je1}) becomes:
\begin{equation}
\label{rr2pme}
	{\cal J}^* = 
	\left( \begin{array}{cccccc} 
	     - ( \nabla_x d_1 )^T \cdot {\cal J} \cdot \nabla_x d_1 \;\: & 
	\;\: - ( \nabla_x d_2 )^T \cdot {\cal J} \cdot \nabla_x d_1 \;\: & 
		\vline & 0 & \ldots & 0 \\
	     - ( \nabla_x d_1 )^T \cdot {\cal J} \cdot \nabla_x d_2 \;\: & 
	\;\: - ( \nabla_x d_2 )^T \cdot {\cal J} \cdot \nabla_x d_2 \;\: & 
		\vline & 0 & \ldots & 0 \\  \hline
	0 & 0 & \vline & 0 & \ldots & 0 \\
	\vdots & \vdots & \vline & \vdots & \mbox{} & \vdots \\
	0 & 0 & \vline & 0 & \ldots & 0 \\
	\end{array} \right)
\end{equation}
But now recall that for any two functions $f(x)$ and $g(x)$ it is:
\begin{equation}
\label{rr2cm}
	\{ f(x), g(x) \}_{{\cal J}(x)} = ( \nabla_x f )^T \cdot {\cal J} \cdot \nabla_x g = 
	\sum _{i,j=1}^n (\partial _{x_i}f) J_{ij} (\partial_{x_j}g)
\end{equation}
Thus in (\ref{rr2cm}) expression $\{ f(x), g(x) \}_{{\cal J}(x)}$ denotes the Poisson bracket of $f(x)$ and $g(x)$ in the sense determined by ${\cal J}(x)$. According to (\ref{rr2pme}) and 
(\ref{rr2cm}) we arrive at:
\[ 
J^*_{11}= \{ d_1(x),d_1(x) \} _{{\cal J}}=0 \:\; , \:\;\:\; 
J^*_{22}= \{ d_2(x),d_2(x) \}_{{\cal J}}=0
\] 
and $J^*_{12}= \{ d_1(x),d_2(x) \}_{{\cal J}}= - J^*_{21}$. Consistently we obtain by construction that matrix ${\cal J}^*$ is skew-symmetric. Moreover, since Rank(${\cal J}$)$\, =2$ and Rank($M$)$\, =n$ everywhere in $\Omega$ by hypothesis, matrix ${\cal J}^*$ is congruent on $\mathbb{R}$ with ${\cal J}$ and then it is also Rank(${\cal J}^*$)$\, =2$ everywhere in $\Omega ^* = y( \Omega )$. Accordingly it is $J^*_{12}(y) \neq 0$ everywhere in $\Omega ^*$. This implies that in order to fulfill the Darboux reduction we only need to perform the time reparametrization $\mbox{\rm d} \tau = \eta (y) \mbox{\rm d} t$ as detailed in (\ref{jr2darbntt}), which is well defined everywhere because now
\[
	\eta (y) = \left( \left. \{ d_1(x), d_2(x) \}_{\cal J} \right) \right| _{x(y)} = 
	\left[ \left. (\nabla_x d_1(x))^T \cdot {\cal J}(x) \cdot ( \nabla_x d_2(x)) \right] 	\right| _{x(y)} = J^*_{12}(y)
\]
is $C^{\infty}(\Omega ^*)$ and does not vanish in $\Omega ^*$. Evidently, this time reparametrization transforms the structure matrix ${\cal J}^*$ into the Darboux canonical one, thus completing the global reduction. In order to conclude the proof, it is only required to demonstrate that transformation (\ref{jr2tdf}) is a global diffeomorphism in $\Omega$. This is actually a consequence of several facts: the change of coordinates (\ref{jr2tdf}) is a function globally onto (since $\Omega ^* = y( \Omega )$ by definition) and by hypothesis one-to-one in $\Omega$. Consequently, (\ref{jr2tdf}) is a global bijection and the inverse function of (\ref{jr2tdf}) exists everywhere and is unique. Moreover, both the transformation (\ref{jr2tdf}) and its inverse are globally differentiable (and therefore continuous) since the functions $(d_1(x), d_2(x), D_3(x), \ldots D_n(x))$ are $C^{\infty}( \Omega )$ and $\mid \! M \! \mid \neq 0$ in all points of $\Omega$, as indicated in (\ref{rr2j1}). \hfill \mbox{\bf {\small Q.E.D.}}

\mbox{}

It is worth recalling that the contribution given in Theorem 5.1 generalizes the local result ensured by Darboux' theorem for the case of arbitrary dimension and rank two. Additionally, since Theorem 5.1 is a purely $n$-dimensional construction, it provides a kind of development not very frequent in the literature, apart from some exceptions 
\cite{bsn1,bsnf,byv2,byv4,pyj1}. Moreover, the generality of the result presented is clear from the fact that some Darboux reductions reported in the literature \cite{dddh,bs2,bs4,bs3,bsnf} actually become particular cases of the algorithm developed in Theorem 5.1.

\mbox{}

\mbox{}

\end{document}